# On the possible discovery of precessional effects in ancient astronomy.

Giulio Magli
Dipartimento di Matematica del Politecnico di Milano
P.le Leonardo da Vinci 32, 20133 Milano, Italy.

Abstract:
The possible discovery, by ancient astronomers, of the slow drift in the stellar configurations due to the precessional movement of the earth's axis has been proposed several times and, in particular, has been considered as the fundamental key in the interpretation of myths by Ugo de Santillana and Ertha Von Dechend. Finding clear proofs that this discovery actually occurred would, therefore, be of relevant importance in a wide inter-disciplinary area of sciences which includes both social-historical and archaeo-astronomical research. In the present paper the possible discovery of astronomical effects induced by precession - such as the shift in the declination of the heliacal raising of bright stars or the so called precession of the equinoxes - is analysed for various ancient cultures in the world. Although definitive evidence of the discovery is still lacking, the quantity of hints emerging from the general picture is impressive and stimulating in view of further research.

Plan of the paper:
1  Introduction

2 Astronomical data

2.1 Babylonian culture
2.2 The indo-savrastati culture
2.3 Egypt : Middle and New Kingdom astronomical data
2.4 The maya

3  Astronomical alignments

3.1 Egypt : orientation of temples
3.2 Egypt : orientation of pyramids
3.3 Malta
3.4 Majorca
3.5 Sardinia
3.6 The Medicine wheels
3.7 Teutihuacan and the "17 degree" family

4 Post-discovery hints

4.1 The cult of Mithras
4.2 The Gundestrup cauldron

5 Concluding remarks
Acknowledgements
References

# 1 Introduction

The earth rotates around its axis in 24 hours, and the earth's axis rotates around the axis orthogonal to the ecliptic, describing a cone. Thus, the motion of the earth is similar to that of a top: the earth precedes. The period of this movement is extremely long with respect to human life, since the axis completes a cycle in 25776 years.

Precession has a very important consequence on long-term naked eye astronomy. First of all, the prolongation of the earth axis on the celestial sphere defines astronomical north. The direction in which astronomical north points – possibly indicating a star, thereby a pole star – changes therefore continuously in time. Today's pole star (our Polaris) will resign to be the pole star in a few centuries, and all the stars which lie close to the circle described by the pole (actually not exactly a closed circle, due to perturbations) will become "pole stars" one time each precessional cycle. For instance, in Palaeolithic times, the north pole crossed the Milky Way and the Pole star in 15000 BC was Delta-Cygnus. The north-pole sky was therefore completely different from ours; it was probably depicted in a fresco of the famous Lascaux grotto (Rappenglueck 1998).

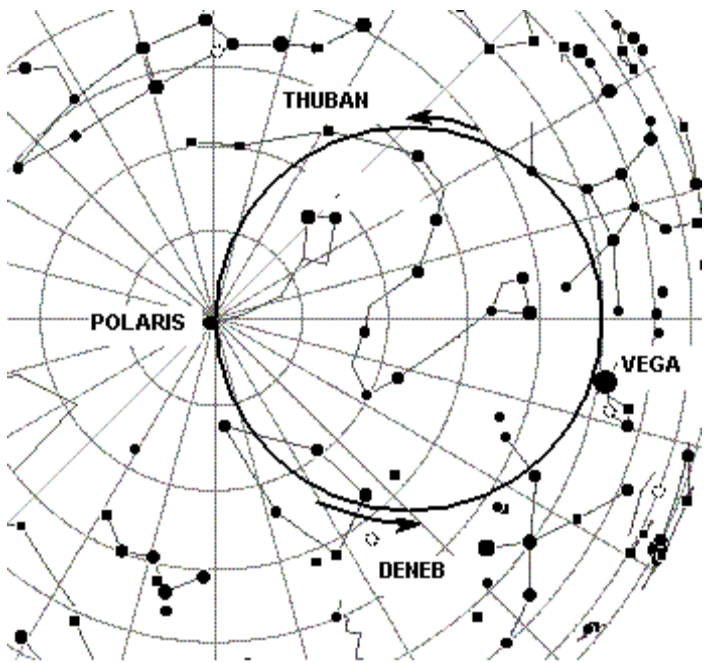

Fig. 1 The circle described in the sky by the north celestial pole during a precessional cycle

Although at best visualized using the movement of the north pole (or actually of the south pole, choice of north is only due to the latitude of the present author at the moment of writing) precessional effects act on all stars. For instance, precession slowly moves the rising point of non-circumpolar stars and slowly moves the culmination of stars: it follows, that the whole visible sky in a given point at a given time depends on the "precessional moment". As an example, one can consider the constellation of the group of stars Crux-Centaurus at the latitudes of the Mediterranean sea (the "South Cross" constellation was "isolated" as a standing constellation only the 16 century AD). This asterism was quite important for people living at that latitudes in very ancient times, as the research by Michael Hoskin and collaborators on megalithic structures in the Baleary Islands and in Malta has shown (we shall come back on this later). However, the Crux-Centaurus group became lower and lower at the horizon during the centuries, and today it "culminates below the south horizon" due to precession and it is, therefore, invisible (it will be back here only in 12000 AD).
The question now arises, when was the discovery of precession actually achieved. "Standard" scientific point of view states the following:

1) Precession was first discovered around 128 BC by Hipparchus of Rhodes
2) Precession was never discovered in pre-Columbian cultures, in other words it was not known in the Americas before Columbus

In spite of this, the very opposite idea that actually *all* archaic civilizations discovered precession very early has been around for a long time and was stated in a authoritative way by Ugo de Santillana and Ertha Von Dechend (1983) in their famous book *Hamlet's Mill*. Santillana and Von Dechend actually put the discovery of precession as the common root of most, if not all, cosmological myths around the world.
Altough being an extremely interesting and worth reading book, the *Hamlet's Mill* cannot be of any help when discussing the *basic* issue of the discovery of precession, since all the "proofs" recorded in it cannot be considered as true scientific proofs. Indeed the authors report an (albeit impressive) amount of occurrences of similar images, same numbers, similar situations in several cosmological myths around the world. Altough it is well known that the myth has actually been used also as a technical language, without a independent, rigorous verification it is of course impossible to accept "images" and "numbers" as "proofs".
The aim of the present paper is to analyse in a systematic way the hints that we really have pointing to the discovery of precession "before" Hipparchus in ancient cultures. To the best of my knowledge this is the first time that such a systematic attempt is made, and it is my hope that the work can contribute to stimulate further research in this field.

## 2 Astronomical data

It is nearly impossible for a naked-eye astronomer (even if very old and expert) to discover precession in the course of his own life using only his own observations, due to the extremely slow nature of the phenomenon with respect to the length of human life. It is, however, sufficient to have astronomical data collected during - say - two or three centuries, and *to trust in them,* to become aware that "something is happening" in the sky with a very low, but measurable, velocity (this is exactly what happened to Hipparchus: he collected a great quantity of astronomical data over more than 800 celestial objects coming from the Alexandria observatory and based his discovery on such data). I shall, thus, be concerned here with the "discovery that something is happening". This means, that I am not speaking about the possible discovery of the actual mechanism and/or of the length of the precessional cycle (although this discovery is not a priori excluded) but rather, of the observation of a discrepancies in specific sets of data: from now on I will call them *precessional effects*. Typical examples may include the observation of the "Precessional Era" namely, the fact that the sun at the spring equinox rises "in different places" within a constellation and finally "changes constellation" every 2000 years, or the observation of the changes of declination of heliacal rising, or height of transit, of a star.

**2.1 Babylonian culture**

In Mesopotamia, astronomers have been collecting astronomical data on argilla tablets for thousands of years. Their records contain observations which are more precise than one minute of arc. Since it is nearly impossible to obtain such an accuracy at naked eye, it was probably reached with the first spyglasses ever invented (Pettinato 1998).

One example of a Babylonian star catalogue is the famous *Mul-apin*. Probably written around 1000 BC, it contains astronomical data which can be traced back in time up to 2048 BC. The content includes:

1) A list of 71 celestial objects (constellations, single stars and the five planets) divided in three "courses" (Enlil, Anu ed Ea).
2) A list of heliacal rising of many stars
3) A list of simultaneous rising/settings of couples of stars
4) A list of time delays between the rising of the same stars
5) A list of simultaneous transit/rising of some others couples of stars.

It is difficult to believe, that astronomers possessing data so accurate did not notice the effect of precession, an idea put forward more than one century ago by the so-called Panbabylonists.
As a matter of fact, however, no written record citing the phenomenon explicitly has been discovered so far.

**2.2 The indo-savrastati culture**

The history of the Indian civilization has been plagued by the fool and anti-historical idea of the so called Arian invasion. The basis of this idea was that civilization was brought in India by indo-European people, the Arians, around 1000 BC. After the discovery of the 2500 BC towns of Harappa and Moenjo-daro, the Arians started to be considered warriors and invaders, but the idea remained, that the fundamental books of the Hindu religion, the *Veda*, where conceived after this invasion. Today, however, we finally do know that the Arians simply never existed and that the Indian civilization (traditionally associated with the sites of Harappa and Moenjo-daro, but actually much more spread than the area individuated by these two cities) developed in between two rivers, the Indo and the Savrastati river (Feuerstein, Kak, and Frawley 1995). The *Veda* contain explicit reference to the latter river, which however drought around 1900 BC, and thus the books (actually memo-books learned by memory by Brahmins) are at least as old as that period.
Together with this new approach to the *Veda*, in recent years a new approach to what we can now call Vedic astronomy emerged (Kak 2000).
In Vedic astronomy a fundamental role is played by the five visible planets, the sun, and the moon, identified with seven fundamental deities. However, to keep track of the motion of them, 27 astronomical objects were used, the *Naksatras*, asterisms/constellations used to divide the ecliptic in equal parts, in each one the sun is "resting" about 13 and 1/3 days. *Naksatras* occur in ordered lists. For instance, one reads (with modern names) Pleiades, *alfa-tauri* (Aldebaran), *beta-tauri, gamma-gemini, beta-gemini* (Pollux), *delta-cancri*, Hidra, Regolus, and so on. Interestingly enough, lists of Naksatras belonging to different periods contain the same objects but begin in different points. The starting point is individuated by the sun at the spring equinox, and this means that Vedic astronomers were almost certainly aware that the Sun was "changing Naksatra" with a velocity of more than one Naksatra per millennium (25776/27).

**2.3 Egypt : Middle and New Kingdom astronomical data**

The study of the ancient astronomy in Egypt has been plagued for many years by the influence of the most important scholar in the field, Otto Neugebauer, who stated in several occasions assertions like this: "Egypt did not contribute to the history of mathematical astronomy" (see e.g. Neugebauer 1969, 1976).
Curiously enough, it just suffices to read the information contained in the monumental books by the very same Otto Neugebauer and by Richard Parker on ancient Egyptian astronomical texts (1964), to see how such an assertion is far from being even a little bit true.

Another serious problem generated by the bad influence exerted by Neugebauer is the idea that astronomy was not present in the Pyramid Age, and in fact the Neugebauer-Parker book begins with the Middle Kingdom (we shall see later that also this assertion is clearly false).

Much of the confusion arises from the fact that we do not have any Egyptian text of explicit astronomical nature coming from pharaonic times, a thing that, in my opinion, is probably due to the fact that such papyri simply were not part of the funerary items, and practically only such items are being recovered. In any case, it is obvious that Egyptian astronomers did actually keep track of many astronomical events. This is readable from those "astronomical texts" which were used in funerary contexts such as those depicted in Middle Kingdom coffins and in many New Kingdom tombs, such as the famous tomb of Senmut, architect of the Queen Hatshepsut, and the ramesside tombs of the King Valley.

In the Middle Kingdom, the so-called decanal lists were used. Decans were 36 stars (or groups of stars) whose heliacal rising (the day of the first raising before dawn after a period of conjunction with the sun, i.e. invisibility) occurred in subsequent "weeks" (Egyptian week was made out of 10 days). In this way, the 365-days Egyptian calendar was divided in decans (36x10) plus 5 epagomenal days associated to special decans as well.

It was shown by Neugebauer and Parker that possible decans must lie in a band south of the ecliptic (decanal band) but they considered explicit identification of decans to be impossible. This is untrue and, in fact, today we do have a quite clear picture of which stars the decans represented (Belmonte 2001a,b). Decans were used to keep track of time during the night as well. This is proved by the so called Star Clocks, in which hours during the night are counted associating the last hour of the first day with the decan which has heliacal rise in that day. After one "week" the rising of this decan shifted back in time to sign the previous hour, and another decan signals the last hour, and so on for 12 times (of course each hour has a non–fixed length).

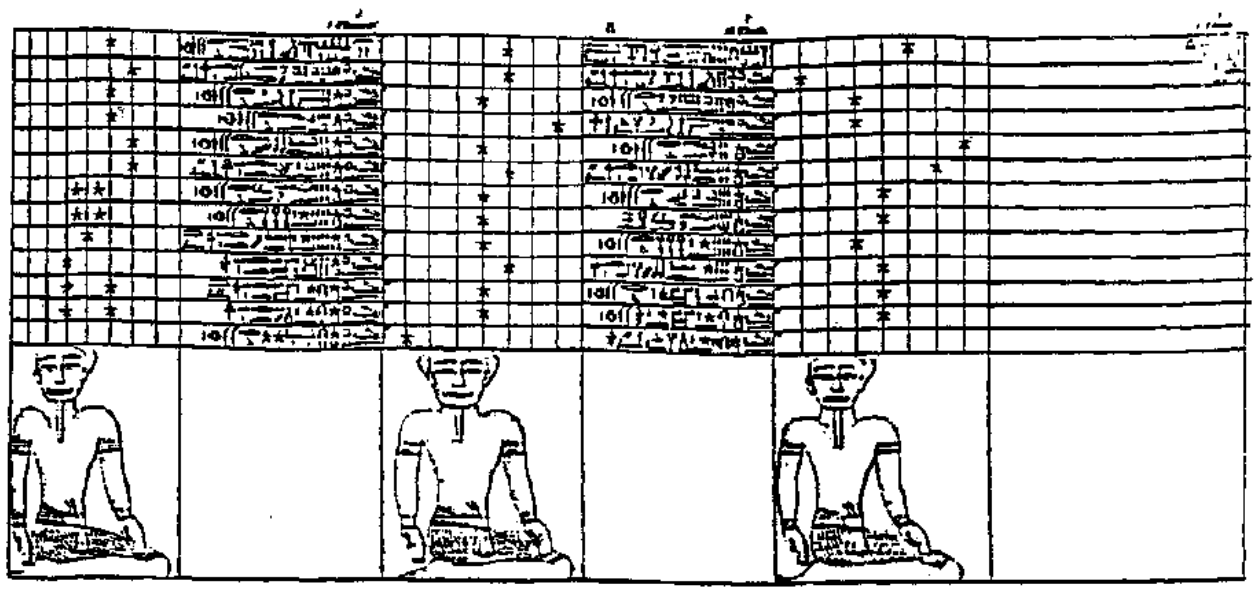

Fig. 2 Examples of ramesside star clocks.

In the New Kingdom stars were observed at the meridian transit rather than at rising, but the way of keeping track of stellar events was similar. This is evident in the so called ramesside star clocks. In a ramesside star clock a men is seen behind a list of 9 columns and 13 arrows. Arrows are associated with hours of the night, columns with parts of the "reference men" and spots sign the transit or position of suitable stars during the night. The framework was changed each 15 days.

I will not enter into further details on the problems of interpretations of such texts. The point I want to stress here is rather that such astronomical devices, although depicted in the tombs (as "guides to the soul during the night") were almost certainly copied from scientific sources (the reader can, if he likes to, add quotation marks to the word "scientific" but I will not do so). In fact, already in the Middle Kingdom Egyptian astronomers were able to keep accurate track of 36 stellar objects taking into account their motion (hour of rising, period of invisibility and so on) and therefore they should have selected such properties from a huge amount of observational data. It is certain that one can discover a precessional effect in the heliacal rising of a star using data accurate to ½ of a degree in – say – three centuries. This led Pogo (1930) and Zaba (1953) to propose that precession was probably discovered very early in Egypt. It is, in addition, worth mentioning that several authors have proposed, in order to explain the curious arrangements of the constellations in the famous round picture of the sky known as the "Dendera Zodiac", that it could contain a reference to the precessional movement of the north pole (see e.g. Trevisan). The "Zodiac" is however dated to the first half of the last century BC, and therefore it has been sculpted a few decades *after* Hipparchus discovery.
Again, we do not have any explicit record which can be associated unambiguously to the discovery of a precessional effect.

### 2.4 The Maya

As is well known, the Maya kept track of astronomical data in a written and extremely accurate way (see e.g. Aveni 2001). Unfortunately, only four maya "codex's" survived to the *autodafe'* to which the bishop of Yucatan, Diego de Landa, condemned all the "heretic books". Such codex's contain data about eclipses, about Venus and about Mercury. Data are so precise (for instance, the Venus table in the Dresda codex is based on tens of years of observations) that the ability of the maya astronomers in taking extremely accurate measures is beyond any doubt. However one cannot discover precession studying the motion of the sun, of the planets and of the moon, and we do not have any record of stellar observations by the Maya (the unique exception possibly being in the so called Paris codex, which is still partially not understood).

# 3 Astronomical alignments

So far, we have looked for written evidences, without much success indeed. There is, however, another possibility to keep track of celestial motions and, consequently, to live astronomical data to successors as an heritage: that of constructing stellar alignments. Following the accuracy of the alignments during a few centuries one can easy discover precessional effects (kindly notice that I am using here an abuse of notation calling "stellar" the alignments pointing to stars *different* from the sun).

### 3.1 Egypt : orientation of temples

The pioneer in the studies of the astronomical orientation of temples in Egypt was Norman Lockyer (1894). In his book, he studied orientation of many temples, but I shall discuss in details here only the case of the two main Theban temples, Karnak and Luxor.
These two temples have a millenary history and were embellished and enlarged several times. In particular, different pharaons in different epochs "added" further galleries in the direction of the main axis of both temples. If one looks at the plan of the Karnak temple, it is clearly seen that the building was always enlarged maintaining strictly the original direction of the main axis. It was shown by Lockyer that this direction is that of the setting sun of the summer solstice. The work of Lockyer was criticized because hills at the horizon prevent the setting sun to penetrate the gallery, and today we actually know that the orientations of temples always took into account *also* the position of the Nile (for instance, to allow for the arrival of processions from the river). Thus, the front of the temple was oriented towards the Nile *and* the setting sun, while observations were performed at the other end of the temple in a chapel which – being in axis with the temple - is oriented to the winter solstice sunrise (Krupp 1983, 1988). In any case, solstice alignment of the temple is certain, and of course, precession does not effect the apparent motion of the sun so that any enlargement was added in the same direction.

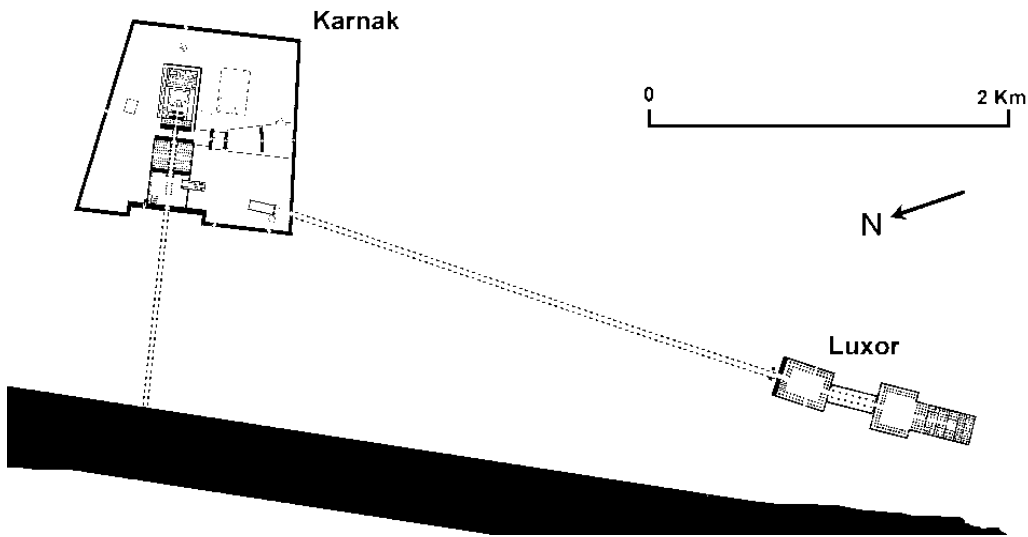

Fig. 3 Plan of Karnak and Luxor temples.

The second main temple of Thebes, today's called Luxor temple, does not exhibit a straight axis. It is oriented roughly east of north and roughly parallel to the Nile, but the axis was *slightly* deviated not less than four times, every time in the occasion of a subsequent enlargement which took place during the centuries. This led Lockier to conclude that the temple was aligned to a star, and that precession caused the slight deviations of the axis in the course of successive enlargements. Unfortunately, although we do have several description of the alignment ceremony of temples to the stars, which was called by the Egyptians *Stretching of the Cord*, we do not have a clear picture of how the ceremony actually took place. For instance, in many cases it is said that the alignment occurred towards the *Mes* constellation, i.e. the Big Dipper which the Egyptian saw as a Bull's Foreleg, but we do not know exactly to which star it was made. It is therefore as yet unclear if the axis of the Luxor temple can really be associated without doubts to the precessional shift of an astronomical event, or not.
Other examples pointing to the same conclusion include the temples at Medinet-Habu and the Isis chapel at Dendera. This building was aligned to the heliacal rising of Sirius in 54 B.C. However, the Chapel was erected on the foundations of a pre-existing ramesside building whose axis was aligned to the same astronomical event about 1250 years before and – therefore – was shifted of about 2.5 degrees (Aubourg & Cauville-Colin 1992)

**3.2 Egypt : orientation of pyramids**

It is very well known that the main pyramids of the fourth dynasty (the main three at
Giza and the two Snefru pyramids at Dashur) were oriented to face to the cardinal points with very high precision. The deviation of the east side from true north is in fact the following (Fig. 4):

(1) Meidum –20' ± 1.0' ;
(2) Bent Pyramid -17.3' ± 0.2' ;
(3) Red Pyramid -8.7' ± 0.2' ;

(4) ' Giza 2 (Khafre) -6.0' ± 0.2' ;
(5) Giza 1 (Khufu) -3.4' ± 0.2' ;
(6) Giza 3 (Menkaure) +12.4' ± 1.0'.

The precision achieved by the pyramid builders is so good that it is absolutely certain that the orientation method used was based on stars and not on the measurement of shadows[1]. The stellar methods which have been proposed in the past, like e.g. the observation of rising and setting of a bright star on an artificial horizon, are not affected by precession. However, as already noticed by Haack (1984), the data strongly point to the existence of a time-dependent font of systematic error and this font is certainly precession. The problem aimed Kate Spence (2000) to propose a method of orientation – "simultaneous transit" – which consists in observing the cord connecting two circumpolar stars, namely Kochab (*beta-UMi*) and Mizar (*zeta-UMa*) when it is orthogonal to the horizon. Due to the precessional motion of earth axis the cord does not identify always the true north: it has a slow movement which brought it from the `left` to the `right` of the pole in the 25 century B.C. Plotting the deviation from north against time, Spence shows that the corresponding straight line fits well with the deviation of the pyramids w.r. to true north if the date of "orientation ceremony" occurred for the Giza 1 pyramid in 2467 BC ±5y (although no written evidence of orientation ceremony exists for the old kingdom pyramids, the "Stretching of the Cord" foundation ceremony is actually already present in the old kingdom stele called "Pietra di Palermo"). If one, in turn, accepts the method as the one effectively used, the plot can be used to calibrate the date of construction of all the fourth dynasty pyramids, which turns out to be somewhat 80 years later than usually accepted.

Further to Spence work, Belmonte (2001c) proposed that the method actually used consisted in measuring alignments between two stars - as Spence proposed - but using a couple of stars (probably Megrez (*delta-UMa*) and Phecda (*gamma-UMa*)) which are not each other opposite to the pole. The pole is thus obtained by elongation of a cord lying below or over it. This looks more natural (at least for modern naked-eye sky-watchers) and reconciles the astronomical chronology with the usually accepted one. However, it should be noted that the astronomical dating of the so called "air shafts" of the Giza 1 pyramid (Trimble 1964, Badawy 1964, Bauval 1993) points rather to support Spence's later chronology.

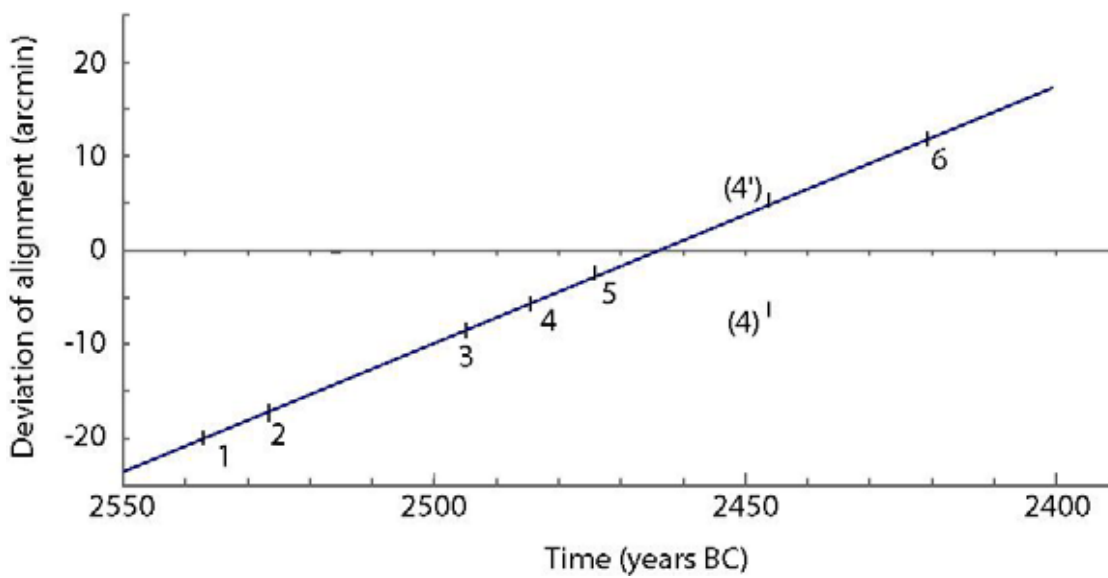

Fig. 4

---

[1] Recently, the French mission directed by M. Valloggia has determined the orientation of the pyramid at Abu Roash (Mathieu 2001), probably constructed by Djedefre who ruled between Khufu and Khafre, to be -48.7', but this error is so out of stream with respect to the others that it points to a different, perhaps solar, orientation ceremony.

As is clearly seen from Fig. 4, the orientation of the Giza 2 pyramid fits in Spence's calibration line if and only if the corresponding point is "lifted up" vertically in the positive region, that is, if (4)' is "lifted" up to (4'). To obtain this, Spence speculates that the orientation of the Giza 2 pyramid was carried out in the opposite season (summer instead of winter) with respect to the others (also in the Belmonte proposal the problem arises and has to be solved assuming a special procedure for the orientation of the Giza 2 pyramid). I tend to think that a ceremony of religious nature, such as the orientation of a giant king's tomb, could not occur scattered in time but should rather occur at a fixed date dictated by astronomical counting, such as e.g. those rituals connected with the Sirius cycle, and I have, therefore, proposed that the error in the orientation of the "second pyramid" actually shows that it was constructed before Giza 1 (and thus occupies the "natural" point (4) of the figure) or, more precisely, that the two projects were conceived together (it can be shown that this "heretical" idea is not in contrast with any indubitable archaeological evidence: see Magli 2003 for details).

In any case, what is really interesting for us here is that the orientation errors of the pyramids form a set of experimental data from which a precessional effect can be deduced. Whatever the reason can be, the effect is negligible in the case of Giza 1 and Giza 2. However one can speculate, for instance, that the relative orientation of the Giza 3 pyramid could have been compared with that of Giza 2 and therefore the precessional effect leading to an angle of 6+12,4=18,4 ' i.e. about 1/3 of a degree, be observed.

In any case, it is worth noting that the astronomically anchored data coming from Giza (orientation of "air-shafts" and pyramids) together with the many astronomical references which are present in the Pyramid Texts do show beyond any possible doubt that astronomy was present in the Old Kingdom as a fundamental part of thinking (religion and knowledge).

### 3.3 Malta

Strangely enough, the Mediterranean archipelago of Malta (composed by the isles Malta, Gozo and Comino) has a short history which, according to all sources, begins only in the fifth millennium BC when Malta was first colonized by humans (Trump 1991,2002). However, after only 1500 years, with the beginning of the so called temples period (3500-2500 BC) Malta civilization became the first to construct megalithic buildings – i.e., not tombs - on the whole earth.

In the megalithic phase more than 40 temples were constructed. Actually the world "temple" should be put in quotation marks because it is far from being clear which was the real function of the building. However traces of a worship for a "mother goddess" deity are evident. The temples are composed by buildings (up to three, corresponding to subsequent phases and numerated accordingly) which all have an external masonry of ovoid shape while the internal plan is composed by a subsequent series of "lobs" constructed along the same axis and ending with an "apse". The internal "lobs" probably recall the shape of the "Mother Goddess".

The best preserved temples are *Ggantija*, the place of the giants, in Gozo, and *Hagar Qim*, *Mnajdra* and *Tarxien* in Malta.

The temples show a clear interest of the builders for celestial phenomena. This interest is evident in Mnajdra II, which is a solar calendar built in stone: the axis is aligned due east, and the "altar" stones are put in such a way that one can keep track of the yearly movement of the sun from the far left to the far right of the "apse".

All other Malta temples have axes oriented south of east, and their orientation is too far south to be related to the sun and the moon (i.e. it is south of the southern moon major standstill rising azimuth). Due to the work by Michael Hoskin and collaborators and by Klaus Albrecht however, today we have a quite clear interpretation of these orientations. As a key example, I will discuss Ggantjia.

The two temples of Ggantjia correspond to two subsequent phases and the second one is oriented further south with respect to the first. Both exhibit a solar orientation in the left altar, which is oriented to winter solstice sunrise (Albrecht 2001) and both exhibit a stellar orientation in the main axis, which is oriented towards the rising of the asterism composed by the South Cross and the two bright stars of Centaurus (remember that it is only from a few centuries that the South Cross has been identified formally as a constellation) (Hoskin 2001).

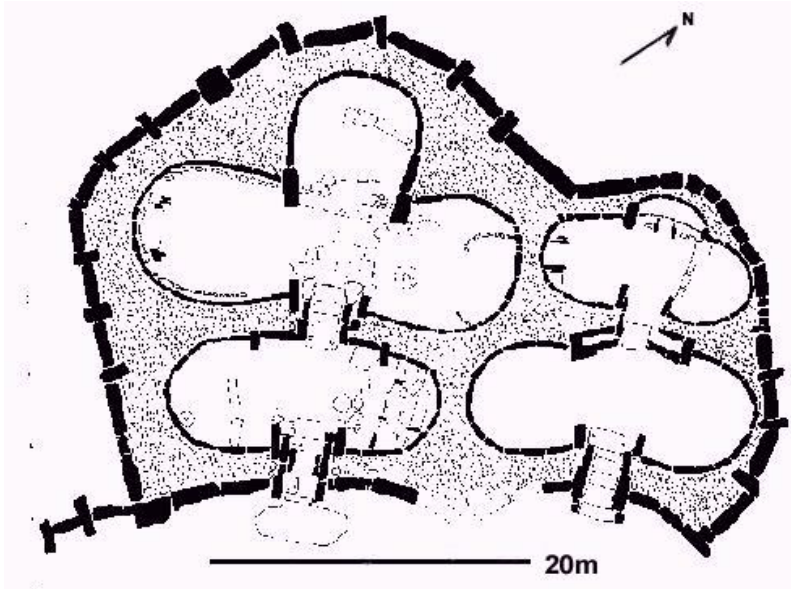

Fig. 5 Plan of the Ggantija temples

Ggantija is thus, in my opinion, a quite clear example in which the builders were interested *both* in solar and stellar orientations. The problem for them was, of course, that while the solar direction was to remain accurate for centuries and centuries, the stellar one was changing, due to precession. It is therefore strongly tempting to conclude that they were *obliged* to construct the second temple due to the movement further south of the raising of the Crux-Centaurus asterism.

**3.4 Majorca**

The isles of Minorca and Majorca were, about one thousand years after Malta and thus during the Bronze Age, inhabited by megalithic sky watchers. The so called sanctuaries of the two islands, including the famous Minorcan *Taulas,* megalithic structures composed by two monoliths disposed a s a giant "T", were oriented to the rising of the same asterisms mentioned before, composed by the South Cross and the two bright stars of Centaurus (Hoskin 2001).
We are interested here especially in one of the sanctuaries, called *Son Mas*, in Majorca.
When the Hoskin group studied the site, it became clear that it was oriented to the low arc in the southern sky that the asterism Crux-Centaurs was following at the end of a valley, in about 2000 BC. However the lower part of this asterism would have become invisible, due to precession, in about 1700 BC Therefore, if the site was really connected with astronomical observations, it should have been abandoned around that date.
Hoskin was not, at that time, aware that a team headed by Mark Van Strydonck of the Belgian Royal Institute of Cultural Heritage was carbon-dating samples from the same site, and was actually wondering why the site was abandoned exactly in that period!

This is thus a very interesting example of the way in which Archaeo-astronomy can act as a predictive science. What is especially interesting for us here is, of course, that it is clear that an astronomical alignment showed that "something was happening" in the southern sky in Majorca and induced the people to abandon the site.

**3.5 Sardinia**

In the Italian island of Sardinia one can visit thousands of *nuraghes*, huge buildings composed by one or more cyclopean towers which were constructed in the period 1800-1000 BC. Altough interpreted by many archaeologists as buildings having a defensive purpose, this hypothesis has never been proved and, as a matter of fact, the incredible number, the smallness of the interior chambers, and the lack of historical proofs of extended internal conflicts and/or external enemies rather point to a symbolic, religious origin for the construction of nuraghes. This idea has been recently supported by the proof of their astronomical orientation, obtained by Zedda and Belmonte (2004).
Summarizing, the results obtained by these authors in a wide sample composed by 272 "simple" (i.e. one tower) Nuraghes and 180 "complex" (i.e. conjunct of different monuments including a central tower) Nuraghes are the following. Orientation is always south of east, and there are three clear peaks, one at the winter solstice, the second at the major southern lunar standstill, and the third, and most populated, pointing to the raising of the Crux-Centaurus group (thus suggesting mutual influences with the Baleary islands). What concerns us here, is the discovery by Zedda and Belmonte that the latter peak *moves towards south* from a declination of about –43 degrees to –45 ½ degrees when one considers separately "simple" nuraghes (presumably built earlier) and "complex" nuraghes (constructed later). This movement corresponds very well to the precessional shift of *alpha-centauri* over the period 1500-1000 BC.
There was, therefore, a clear possibility for the Nuraghes builders to observe a precessional effect.

**3.5 The Medicine wheels**

The so called *Medicine Wheels* are stone monuments composed by a central cairn of stones connected by radial rows to an external circle and other cairns. Most wheels are in Alberta, Canada, but the most famous of them, the Big Horn wheel, lies near the Medicine Mountain in Wyoming and the name of the family comes from this wheel.
 There are several typologies of Medicine Wheels, but some of them have been indubitably linked to astronomical observations. The first to be identified with an astronomical observatory is exactly the Big Horn one. A Solar Physicist, John Eddy, recognized that the small cairns which are distributed on the external circle of the wheel serve as astronomical outpost for many alignments. The alignments recognized by Eddy are at the summer solstice and at the heliacal rising of Aldebaran, Rigel and Sirius (Eddy 1974, 1977). The window of validity of such alignments (which is of the order of three centuries, due to precession) holds for the last tree centuries, and indeed independent archaeological data give to the Big Horn an age of 250 years.
Eddy's interpretation received a wonderful confirmation after the archaeological study of another wheel, the Moose Mountain Medicine wheel. This wheel in fact has alignments towards the same targets, but the window of validity is completely different, and lies around the last centuries BC When archaeologists Tom and Alice Kecoe obtained c-14 datable samples from the site, they were able to confirm the "astronomically predicted" age of the monument, showing a constant interest of the (at yet unknown) wheel builders for the same astronomical objects in the courses of two millennia: another wonderful example of Archaeoastronomy as a predictive science (actually the astronomical tradition of the wheels is much older than this: the Majorville Wheel in Alberta was used for solar observations already in 2500 BC).

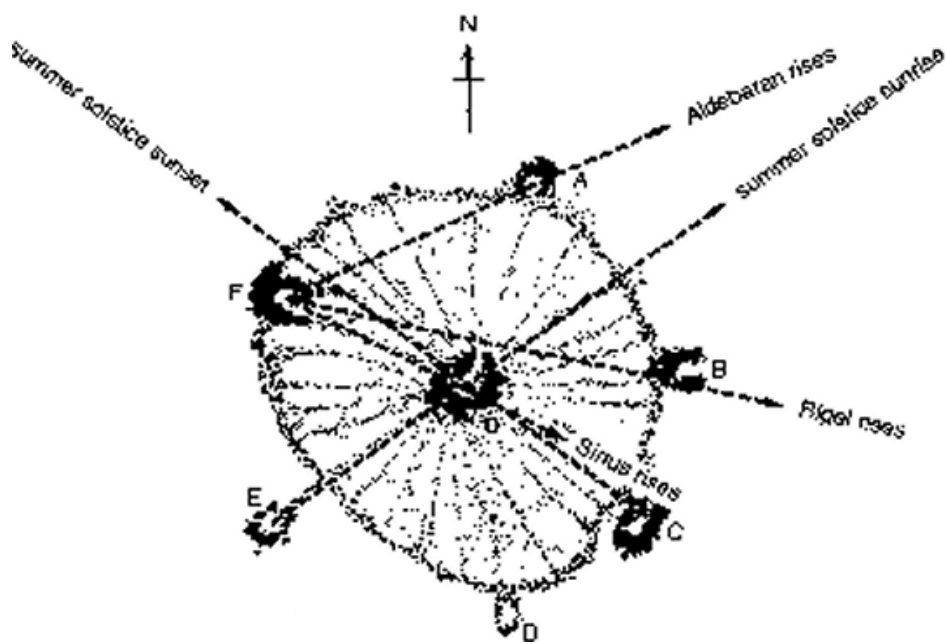

Fig. 6 The astronomical alignments of the Big Horn Medicine Wheel discovered by Eddy.

The astronomical purpose of the Medicine Wheels has been strongly criticized in the past, especially due to the fact that the accuracy of the alignments is poor. However, the very same fact that the builders where interested to the phenomena discovered by Eddy (independently from the precision they wanted to obtain in measuring them) is proved beyond doubts (see Aveni 2003 for a recent discussion).

What is especially interesting for us here is the missing "D" alignment, which was individuated by Robinson (1980) both at Big Horn and at Moose Mountain. Robinson discovered this direction to be aligned with the rising of Fomalhaut, a star of the constellation Piscis Austrinus. The window of validity of this alignment is however *shifted in time* of some centuries with respect to Eddy's estimates for Moose Mountain. This looks strange, but the radial line of stones is curved along its length. It looks like that the line was originally pointing more westerly and was then curved in order to follow the precessional shift of the point of rising of the star.

Thus the Moose Mountain Medicine Wheel strongly candidates as a place where tenacious astronomers observed a precessional effect.

**3.6 Teutihuacan and the "17 degree" family**

While, as we have seen, we have a very clear picture of the way in which the Maya recorded their astronomical observations, the same cannot be said of others Mesoamerica cultures. We practically do not know anything about the astronomy of the so called mother culture of Mesoamerica, the Olmecs, and we do not have written records coming from the most important culture of the Mexico valley, which flourished during the pre-classic maya period, roughly between the second and the sixth century AC, and which influenced all subsequent civilization in central Mexico: Teutihuacan. Teutihuacan lies not far from Mexico city, and it is a huge town which, at the moment of maximum urbanization, should have reached more than 125.000 inhabitants. The city was planned under a rigid project which aimed to *replicate* the landscape. This is evident from the fact that the two main buildings, the so called Sun Pyramid and Moon Pyramid (these are later denominations, no connection with sun and moon has never been proved) are disposed in such a way to be a "copy", an image, of the two mountains which lie respectively on the back, the Cerro Gordo and the Cerro

Patlachique. The town was planned and carefully constructed on a "cardinal grid" based on two axes, a "T-nord" axis oriented 15.5 degree east of north, and a "T-east" axis oriented 16.5 degrees south of east. Teutihuacan "cardinal directions" are thus rotated with respect to the "true" cardinal directions and tilted of one further degree each other.

Astronomy plays here a fundamental role, since the most reasonable explanation is the following. The T-east orientation is a solar orientation. It is too close to east to signal any special event in the motion of the sun at the horizon (solstice and days of zenit passage) however the sun sets at T-west on 13 august and 29 April, and these two dates are separated by 260 days. It is well known that the so called sacred calendar of Mesoamerica (well documented in the Maya, but probably coming from the very early civilization and codified around 400 BC) was composed by 260 days. The origin should be the passage of the sun at zenit, which of course depends on latitude and occurred in those two dates at the latitude of the pre-classic site of Izapa (see Aveni 2001 for a complete discussion). Thus, the T-east orientation was probably a reminder for the sacred calendar of solar origin. What is especially interesting for us here is however the T-nord orientation because it is almost certainly a stellar one.

The axis orthogonal to T-nord (which, just as a remind, is *not* parallel to T-east) is individuated by an accurate alignment between two so called pecked crosses, pecked symbols incised on the ground, one on a hill at the west horizon and the other one in the centre of the town. This alignment points to the setting of the Pleiades around 100-400 AD, and this asterism had heliacal rising approximately in the same day of the zenit passage of the sun (18 May) and culminated near the zenit as well (Dow 1967).

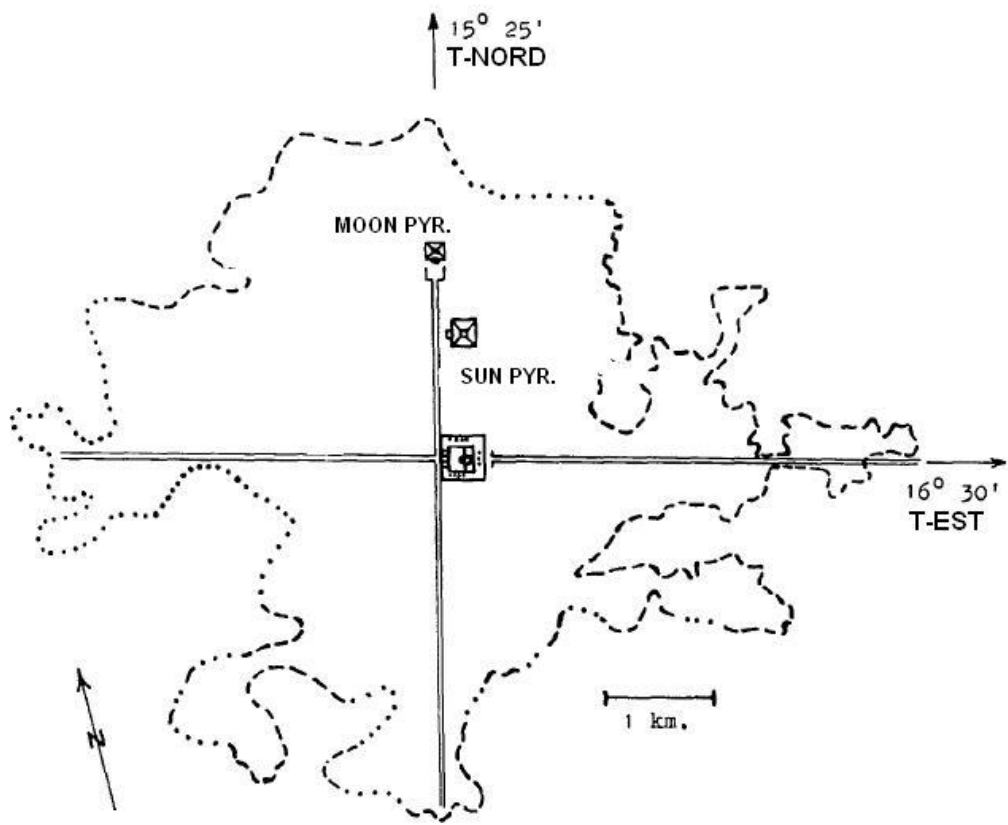

Fig. 7 Map of Teutihuacan

Teutihuacan collapsed a couple of centuries thereafter, and it is therefore unlikely that Teutihuacan astronomers were able to realize that the stellar alignment was becoming less and less accurate due to precession. What is of special interest for us here is what has been called after Aveni and Gibbs the *17 Degree Family* (Aveni and Gibbs 1976).

The family comprises several archaeological sites in central Mexico. All such sites exhibit the same (or very near to) T-nord orientation. The family includes, among others, the first phase of the giant Cholula pyramid, the Toltec temple of Tula and the pyramids of Tenayuca and Tepotzteco, and therefore it comprises buildings which have been constructed several centuries after 400 AD. It is thus clear that the T-nord orientation did not indicate the rising of the Pleiades any more. The question obviously arises, if the architects were aware that they were orienting buildings to a stellar direction which was no more effective for some reasons, and in this case, if they asked themselves for the reasons, or simply if they were doing so "in memory" of the past glory of Teutihuacan without even knowing which was the original meaning of the direction. A plot of the precessional movement of the Pleiades against the supposed date of construction of the buildings taking into account their different latitudes would certainly be of help in assessing this point.

# 4 Post-discovery hints

**4.1 The cult of Mithras**

The facts which I have exposed in the previous section strongly point, in my view, in the direction of showing that precessional effects were actually discovered using astronomical alignments. However, the problem remains, why we do not have *explicit* mention of such effects anywhere. Being a physicist, I like enigmas (i.e. solvable problems) and I do not believe in "mysteries". A tempting explanation for the enigma is, that the discovery was not explicitly stated because the precessional movement was considered a thing to be kept secret, or (better) reserved to a group of "initiated" people. If this is true, it is, of course, a truth very difficult to establish. Our unique possibility is to investigate whether traces of the discovery of precession can be found in cults reserved to initiates, at least in historical times.
Actually, such traces can be found.
As is well known, the cults which where reserved to initiates are called in historic literature *mysteric* cults, one famous example being the so called Eleusi Mysteries in Greece and another being the Mithras Mysteries in the first three centuries AD in the roman empire. Interestingly enough, an extremely intriguing hint pointing to the discovery of precession in ancient times comes exactly from such a cult.
Hipparchus discovers precession around 128 BC, working in Rhodes. About 50 years *later*, Pompey defeats the pirates in the Mediterranean sea, and his legionnaires come into contact with a religion which will rapidly spread in the whole roman empire in the subsequent two centuries, and will be destroyed by the christianisation of the empire: the Mithras cult.
In the Mithras cult the rituals were kept secrets to non-adepts and we do not have any written records describing them. However, several underground "shrines" have been unhearted and studied by the archaeologists, perhaps the most famous of them being the one present in the S. Clemente catacombs in Rome. The iconography of the cult is very well known and is represented (sculpted or painted) in the ending "chapel" of the shrine. We see the god, Mithras, represented as a young men, killing a bull with a sword. The god does not look at the bull. Under the bull, a scorpion hits at the genitals of the bull, and the figures of a dog, a serpent, a crow, a lion and a vessel occur. From the bull's tail some grain ears sprout. Frequently, the zodiacal signs and the planets are represented as well.
The history of modern Mithraic studies is very instructive and almost unbelievable. In 1896 the Belgian scholar Franz Cumont formulated a theory, in which the cult was interpreted as an adaptation of a old iranic cult, *Mithra.* Altough many clear aspects of the Mithras cult were not recognizable in the Mithra cult and, in particular, in the Mithra cult there was no sign of the killing of a bull, the authority of Cumont was so strong that his curious ways of deriving Mithras from the iranic Mithra (for instance, recovering the bull from another iranic myth in which Ahriman, a devil god, kills a bull and Mithra does not appear at all) was accepted *up to 1970!*

This "Cumont dogma" is a "wonderful" example of the risks to which we submit ourselves when the "authority of the giants" (or perhaps supposed giants) is accepted outright.

In any way, finally in 1971 some persons started to bring the dogma to the court, and it became immediately clear that Mithras studies had to be re-started from the very beginning and that the natural point to start with was astronomy. [2]

Since the main personages in the play are Mithras and the Bull, it is clear that the bull has to be identified with Taurus but it is not clear with which constellation has Mithras to be identified. All the astronomical interpretation which have been proposed since 1970 – like e.g. heliacal rising of Taurus - have had serious problem with the identification of Mithras. For instance, one could think to Orion, but Orion is under, and not over, the Bull.

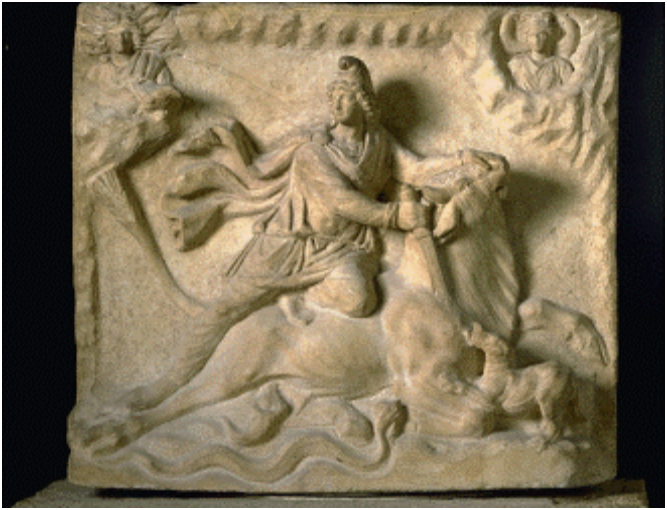

Fig. 8 The Mithras iconography.

Finally, the solution of the puzzle has been given by David Ulansey (Ulansey 1989).
Ulansey observed that *over* Taurus there is Perseus, a constellation identified with a "frigian" warrior already in the 5 century BC. But why the Scorpion? If we send the sky back in time up to the end of the Taurus era, about 2000 BC, we discover that the other equinoctial constellation was Scorpio. The celestial equator crossed at that time Taurus, Canis Major, Hydra (i.e. a serpent) Vessel, Crow and Scorpion (besides a small part of Orion's sword). It remains Lion, which however was the summer solstice constellation at the same epoch. The grain ears from the tail of the Bull give the association with spring equinox.

This is what concerns the interpretation of the Mithras cult: a God who is so strong to be able to change the cosmic order of the motion of the sun with respect to the stars. A very convincing interpretation. However, the interest for us here arises from the way in which Ulansey explains the origin of the Mithras cult.

According lo Ulansey, what happened is (in brief) the following. In 128 BC Hipparchus discovers precession. The discovery rapidly permeates and fits into the symbolic scheme of the stoic philosophy school at Tarso. Since for stoic philosophers, natural forces where manifestations of deities, it was natural for them to introduce a new god responsible for the new movement of the cosmos, a god so strong to be able to move the "fixed" stars. Since Perseus was already venerated at Tarso, the identification followed naturally. Regarding the missing link with the pirates, which are the first Mithras adepts historically documented, Ulansey remarks that they had "contacts with intellectuals" and where used to the stars, being sailors.

---

[2] Actually, already in 1869, the German scholar K. B. Stark noticed strong, clear connections of the iconography with constellations. However Cumont went out to say, that although astronomy could admittedly have played a role in the lower degrees of initiation, the main stream of the high degrees was the iranic tradition on the origin and the end of the world.

I should say that I do strongly believe in Ulansey's interpretation of the Mithras cult but that I am unable to believe in Ulansey's explanation for its origin.

The reason is very simple. Altough doing the best of my efforts, I cannot find even one example in history in which a scientific discovery became a religion. It could eventually became a myth within a religious framework, as in Hamlet's Mill viewpoint, but not the foundation of a cult of a new god. There is also a technical reason for which I cannot believe in Ulansey's interpretation. Let us suppose that a scientific discovery of a mechanism becomes a religion. A religion is usually associated with eschatological thought: we aspect for future event, a future advent of a god, for instance. Therefore, I would rather think that the new religion will be based on the end of the era (Aries to Fish) rather than on the end of the previous one (Taurus to Aries) occurred 2000 years (I repeat, 2000 years) before. Basing on slightly different motivations, this objection has already been raised, and Ulansey's answer is based on the fact that Hipparchus esteem of the precessional velocity was too low (about one degree for century). As a consequence, this led to an estimate of the future change of the precessional era after many centuries (about 800 years) and not at the time it really occurred, actually around the first century AD.

While I consider this as a possible explanation *of the decline* of the Mithras cult (I am not aware of any other making this observation, but it looks natural to me) I do not consider this as a good explanation for its origin, because "time of religion is the time of gods" so there is – usually – no urge for eschatological events to occur.

All in all, I think that the origin of Mithras precessional iconography can be much older than Hipparchus discovery. Once again, these are only speculative statements however. Hopefully new epigraphic or archaeological discoveries might be of help in assessing this interesting point, but at least one archaeological finding already exists.

**4.2 The Gundestrup cauldron**

The so called *Gundestrup Cauldron* is a huge vessel made out of silver plates. Found in Denmark but probably produced by Thracians, it is today exposed in the Copenhagen national Museum and it is the most renewed masterpiece of Celtic art, dated to the first century BC (dating is however only approximate since no physical method is known to date such kind of objects).

The Gundestrup is magnificently decorated with enigmatic images. It shows peculiarities of Celtic religion, like e.g. the god called Cernunnos, but it also shows clear "oriental" influxes (for instance, elephants are represented on one of the plaques). There is still debate about the meaning of the scenes, and what is most debated is the meaning of the representation present in the central plaque. It shows, at the centre, a dying bull with, following a circle around the bull and reading in clockwise direction from upper left, a dog, a warrior, a bear (all with a sort of "fading" incision) and a lizard (or, at least, a 4-legged animal with lizard tail) incised as deep as the bull. A tree branch with leaves is also present, and what seem to be grain ears at the back of the bull.

One can easily solve the exercise of foreseeing which interpretations have been proposed for this image. Of course we have "ritual sacrifice", "ritual fighting with bulls", "ritual fighting between bulls and dogs" and so on (actually the *Corrida* is missing). However, the absence of movement in the scene, the fact that there is no physical connection between the figures, the difference in size between the figures and the presence of the big lizard make these interpretations very doubtful. Recently, the French scholar Paul Verdier (2000) has proposed the idea that the symbolism of the cauldron might be astronomical. For instance, one of the lateral plaques contains two bands separated by a branch. The upper band shows four riders (the solstices) the lower band twelve warriors (the months of the Celtic lunar calendar) while the tree branch is the Milky Way.

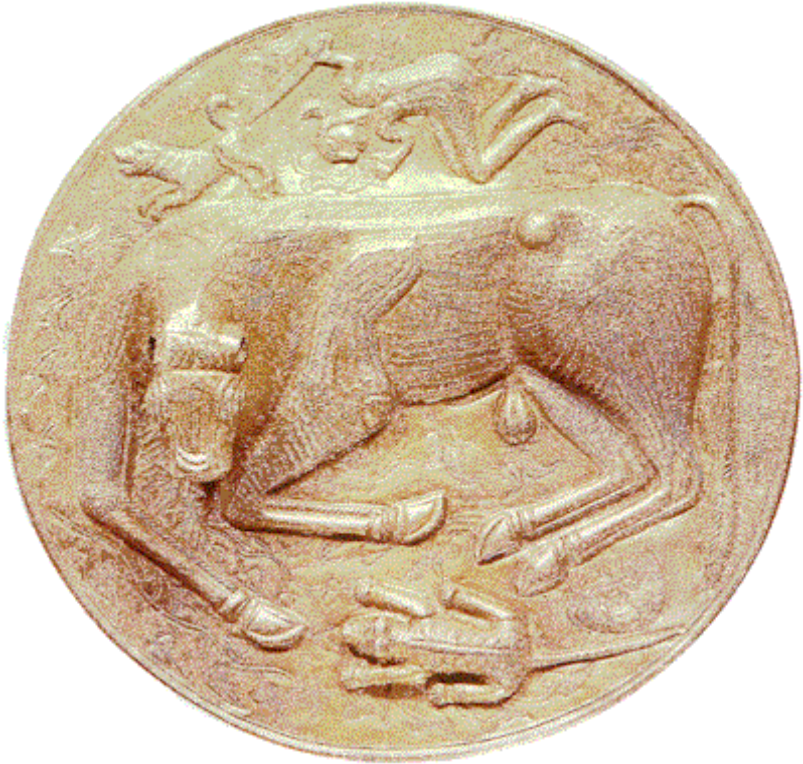

Fig. 9 The central plaque of the Gundestrup cauldron.

According to Verdier, the central plaque should be a representation of the death of the Taurus Era, and the constellations depicted should be Canis Major (the dog) Orion (the warrior) Ursa Minor (the bear) Lacerta (the lizard) and Taurus, the bull. However, it has been noticed by Juan Belmonte (private communication to the author) that this interpretation cannot be correct because the Lacerta constellation is not an "ancient" constellation, since the group of stars forming it have been depicted for the first time as a lizard by the Polish astronomer Johannes Hevelius. Thus, if Verdier's idea of an astronomical interpretation of the Gundestrup is correct, we have to find another "lizard" in the sky of 2000 BC. There are, I think, two candidates for the animal depicted as a lizard, namely the two dragons coming from Babylonian astronomy. One is our Draco, which, sitting near the two Ursae in the northern part of the sky, "follows" Ursa Minor and, at the time of the "death of the bull", was hosting the north celestial pole. The other one, more probable in my view, is the constellation which today we call Cetus. Cetus is today depicted as a whale, but it "sits" under Taurus, exactly as the lizard under the bull in the Gundestrup, and it was identified in ancient times with the dragon-shaped beast Tiamat, the adversary of the Babylonian god Marduk.
If the scene depicted in the Gundestrup is really the death of Taurus era, then the warrior in the scene *might well be Perseus, and not Orion,* since moving clockwise *spiralling* towards the Taurus one actually encounters Perseus, and not Orion. In this case the analogy with the Mithra cult would become striking, also taking into account what seem to be grain ears at the back of the bull.
To the best of my knowledge this is the first time that the central plaque of Gundestrup is proposed to be a representation of the Mithra iconography. Unfortunately however, we do not know the level of astronomical knowledge of the Celtic astronomers, because most of the information we have on them comes from secondary sources, especially (curiously indeed) from the *stoic* Hellenistic writer Posidonio, besides the roman sources like Caesar's writings. However, some primary information is available, like e.g. the *Coligny Calendar*, a lunar calendar written in roman characters but in gallic language. In addition, the lore of astronomy in Bronze Age in North Europe has still to reveal his secrets, as shows the recent discovery of the so called Nebra Disk, a 16 century BC Bronze disk showing 32 stars, a crescent and the sun and probably representing a particular sky in a particular day.

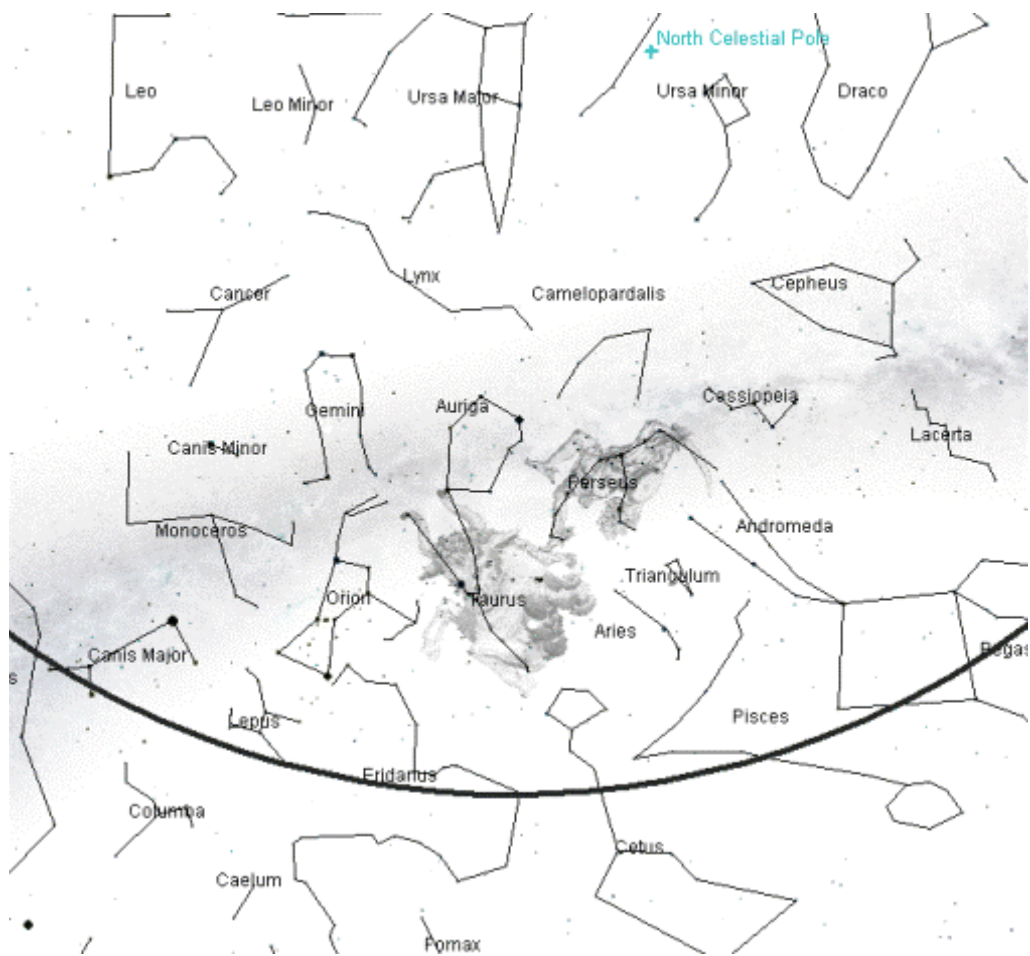

Fig. 10 Portion of the sky at the latitude of Copenhagen, in 2000 BC.

In any case, *if* the astronomical interpretation of the Gundestrup is correct, it is again difficult to believe (at least to me) that also the Celts rapidly filtrated the discovery by Hipparchus, in such a way that an artist of the first century BC decided to represent a precessional event occurred 2000 years before in his masterpiece.

# 5 Concluding remarks

All in all, there is *no* clear, absolute evidence of discovering of precession before Hellenistic times or in pre-Columbian cultures.
There is, however, at least in my view, a clear evidence that some astronomical phenomena, such as the heliacal rising of bright stars or the movement of the equinoctial point trough the constellations, were traced for a sufficient amount of time and with a sufficient precision to lead many ancient astronomers to the discovery that "something was happening" with a very slow velocity with respect to human life.
More research focussed on this issue is certainly needed in many places, but first of all in Egypt. In fact, the problem of the stellar alignment of Egyptian temples should be reconsidered from the very beginning taking into account that the chronology of Egypt is much more clear and accurate than it was in Lockyer times, and controlling the assertions of Lockyer from a *quantitative* point of view (for instance following the subsequent enlargements of the Luxor temple in terms of precessional movement of the stars). Theoretical research is also needed to relate in a secure way decanal lists coming from different centuries.

The need for further research holds true also in Malta, and in all the places which show an interest of the builders for alignments changing in time due to precession.

**Acknowledgements**

Many constructive comments by Juan Belmonte are gratefully acknowledged.